\documentclass[preprint,showpacs,showkeys,aps,prb]{revtex4}
\usepackage[dvips]{graphicx}

\begin{document}

\title{Shockwave Compression and Joule-Thomson Expansion}

\author{
Wm. G. Hoover and Carol G. Hoover                \\
Ruby Valley Research Institute                   \\
Highway Contract 60, Box 601                     \\
Ruby Valley, Nevada 89833                        \\
}                              

\author{
Karl P. Travis                                   \\
Immobilisation Science Laboratory                \\
Department of Materials Science and Engineering  \\
University of Sheffield                          \\
Mappin Street, Sheffield S1 3JD, United Kingdom
}

\date{\today}

\pacs{ 51.30.+i, 62.50.Ef, 47.11.Mn}

\keywords{Joule-Thomson Effect, Shockwaves, Nonequilibrium Molecular Dynamics}

\begin{abstract}
\begin{center}
\noindent

\end{center}

Structurally-stable atomistic one-dimensional shockwaves have long been
simulated by injecting fresh cool particles and extracting old hot particles
at opposite ends of a simulation box.  The resulting shock profiles
demonstrate  tensor temperature, $T_{xx} \neq T_{yy}$ and Maxwell's
delayed response, with stress lagging strainrate and heat flux lagging
temperature gradient.  Here this same geometry, supplemented by a short-ranged
external ``plug'' field, is used to simulate steady Joule-Kelvin throttling flow of hot
dense fluid through a porous plug, producing a dilute and cooler product fluid.
\end{abstract}

\maketitle

\begin{figure}
\vspace{1 cm}
\includegraphics[height=7cm,width=6cm,angle=-90]{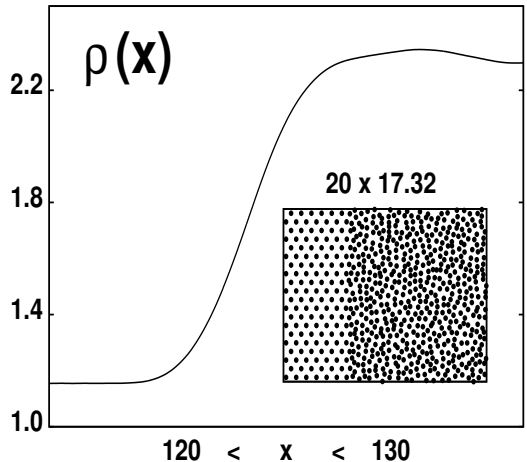}
\caption{
A typical shockwave snapshot.  The motion is left-to-right with
hot fluid exiting at the right boundary.  Density snapshot uses
Lucy's weight function.
}
\end{figure}

\noindent{\it Stationary One-Dimensional Shockwaves}---Shockwaves are arguably
farther from equilibrium than are any other readily available states of a
nonequilibrium fluid\cite{b1,b2,b3,b4,b5,b6,b7,b8,b9}.  In just a few collision
times, or mean free paths, the shock transforms 
cold equilibrium fluid (or solid) into a hot compressed state\cite{b1,b2,b3,b4,b5}.
Laboratory shockwaves at a few terapascals can compress condensed matter as much
as threefold, to densities and pressures far greater than those at the center of
the earth\cite{b6}.  Because the shock transformation is a steady small-scale
continuous process, converting kinetic energy to internal energy without any
external heating, steady-state shockwave structures can be replicated with
computer simulations\cite{b2,b3,b4,b5,b7,b8,b9}.  The inset of Figure 1 shows an interior snapshot
of a typical simulation, with cold particles entering at the left and hot ones
exiting to the right.  The corresponding density profile snapshot using Lucy's
weight function for the spatial averaging\cite{b7,b8,b9,b10,b11} is the smooth
curve. The shockwidth can be estimated from the maximum slope.  It is
just a few atomic spacings.  Steady-state profiles generated in this way are fully
consistent with the transient profiles generated with [1] shrinking periodic boundaries or
[2] headon collisions of two similar blocks of cold
material\cite{b5,b7,b8,b9}.

Both experiments and simulations show that initially-sinusoidal shockfronts
soon become planar.  Steady shockwaves are accurately one-dimensional\cite{b3,b5,b7}.
Accordingly, the mass, momentum, and energy fluxes ( in the $x$ direction, the
propagation direction ) are {\it all} constant in the comoving coordinate frame of Figure 1,
the frame moving with the shockwave\cite{b1} :
$$
\left\{ \ \rho u \ ; \ P_{xx} + \rho u^2 \ ; \ (\rho u)[ \ e + (P_{xx}/\rho) + (u^2/2) \ ]
 + Q_x \ \right\} \ ; \ {\rm All \ Three \ Fluxes \ Constant} \ .
$$ 
Here $\rho(x)$ and $u(x)$ are the mass density and the flow velocity, $P_{xx}(x)$
is the pressure-tensor component in the propagation direction.  $e(x)$ is the internal
energy per unit mass and $Q_x$ is the heat flux vector, measuring the conductive 
flow of heat in the comoving frame.

The cold entrance velocity is $+u_s$ (the ``shock velocity'') and the hot exit velocity
is $+(u_s - u_p)$ (where $u_p$ is the ``particle velocity'') in the hot fluid.  Away
from the shockfront the cold and hot pressure and energy have their
thermodynamic equilibrium values :
$$
P_{xx}(x) \longrightarrow P_{\rm eq} \ ; \ e(x) \longrightarrow e_{\rm eq} \ .
$$
Eliminating $u_s$ and $u_p$ from the three constant-flux equations gives the ``Hugoniot
Equation'' or ``shock adiabat'', $\Delta e = P\Delta v \ $, where $P$ is the mean
pressure, $[ \ P_{\rm cold} + P_{\rm hot} \ ]/2$ , and $\Delta v$ is the overall
change in volume per unit mass, $(1/\rho)_{\rm cold} - (1/\rho)_{\rm hot}$.  Though
there is no {\it external} heating there {\it is} heat flow {\it within} the shockwave structure.
For weak shockwaves it is given by Fourier's Law, $Q_x = -\kappa (dT/dx)$ .

The limiting values of the energy flux divided by the mass flux far from the shockwave are equal :
$$
[ \ e + (P/\rho) + (u_s^2/2) \ ]_{\rm cold}= [ \ e + (P/\rho) + (1/2)(u_s-u_p)^2 \ ]_{\rm hot} \ .
$$
In shockwaves the inflow is supersonic so that the kinetic energy cannot be ignored.
Choosing the initial thermodynamic state along with the particle velocity determines the
shock velocity as well as the pressure and energy of the resulting ``hot'' state. 

\noindent
{\it Joule-Thomson ``Throttling'' Flows}---In the 1850s Joule and Thomson (who became Lord
Kelvin in 1892) collaborated on the design and analysis of experiments seeking to quantify
the ``mechanical equivalent of heat''. {\it The} ``Joule-Thomson'', or
``Joule-Kelvin'', experiment enforced the throttling of a high-pressure gas through a porous plug.  A
detailed description of the evolution of these experiments can be found in Reference 14 .
Within the plug the inlet pressure is reduced to the smaller outlet pressure.  As the flow
rate approaches zero the experiment becomes isenthalpic, where the enthalpy is $E + PV$.
Because there is no external heat flow the work added at the hot high-pressure side less
that extracted on the cold low-pressure side is the energy change :
$$
[ \ e + (P/\rho) \ ]_{\rm high \ P} \stackrel{\rightarrow}{=}
 [ \ e + (P/\rho) \ ]_{\rm low \ P} \ . \ {\rm [ \ Joule-Thomson \ ]}
$$

By contrast to the supersonic shockwave experiment kinetic energy is negligible in the typical
laboratory Joule-Kelvin experiment.  The conductive heat flux (a maximum at the shock front) is
likewise invisible in the throttling experiment, concealed by the irreversible details of the
porous plug.  Otherwise, the geometry and the thermodynamics and the constancy of the fluxes
look identical to the usual one-dimensional shockwave analyses.  In both experiment types there
is necessarily a {\it positive} entropy change within the flow, as is required by the Second
Law of Thermodynamics. 

\noindent
{\it Joule-Thomson Simulations}---The structural similarity of shockwave compression and
Joule-Kelvin expansion experiments suggests the possibility of simulating Joule-Kelvin
flows with molecular dynamics.  Here we validate and illustrate that idea.
Our model must incorporate a computational ``porous plug'' to slow compressed input fluid.
Pores, holes, and confining passageways come to mind.  But a little reflection suggests a
simpler approach---erecting a smooth potential-energy barrier perpendicular to the flow.
This approach is successful.  Apart from the entrance and exit boundaries, the motion is
entirely conservative and Newtonian. The entrance internal energy can be controlled by
adding $y$ displacements and/or Maxwellian velocities $(v_x-u,v_y)$ to particles as they enter.
  
Near the potential plug barrier an anisotropic far-from-equilibrium state results.  The fluid is
first slowed and then accelerated normal to the barrier, with the result that the pressure
and temperature are briefly anisotropic with $P_{xx} > P_{yy}$ and $T_{xx} > T_{yy}$ .
The details of the equilibration involve the same Maxwellian\cite{b13} time delays seen in 
shockwaves.

\begin{figure}
\vspace{1 cm}
\includegraphics[height=9cm,width=6cm,angle=-90]{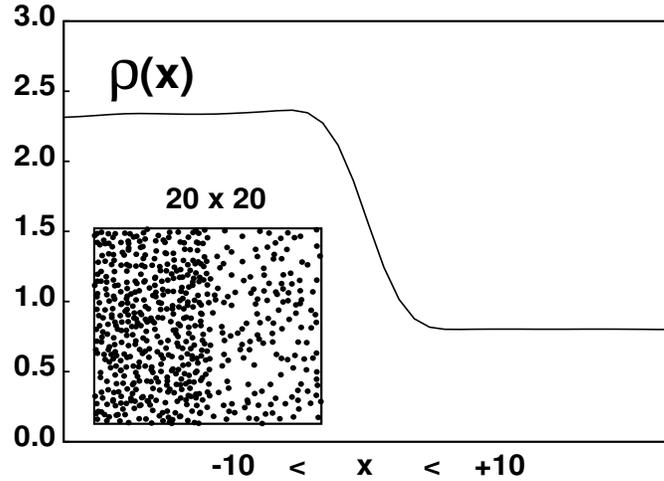}
\caption{
A Joule-Thomson snapshot.  The motion is left-to-right with
cooled fluid exiting at the right boundary.  The density snapshot
uses Lucy's weight function, $(5/12)(1+|x|)[1-(|x|/3)]^3$ .
}
\end{figure}

\begin{figure}
\vspace{1 cm}
\includegraphics[height=15cm,width=10cm,angle=-90]{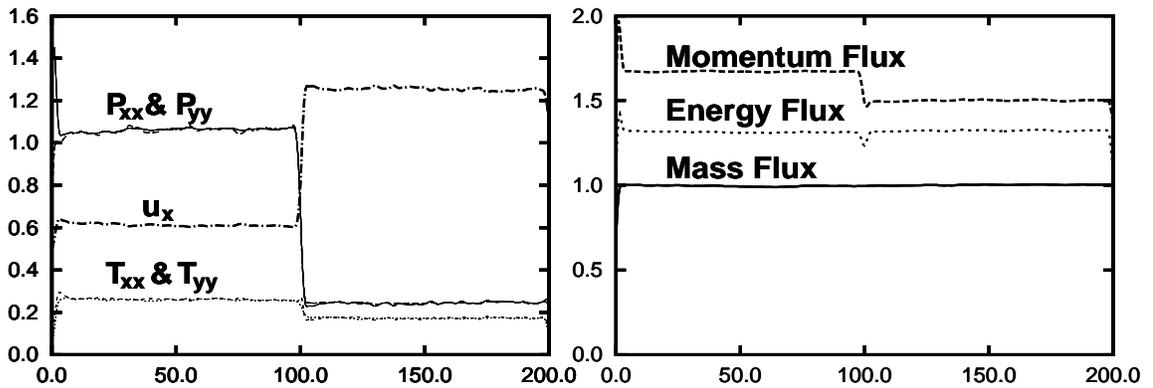}
\caption{
Time-averaged pressure tensor, velocity, and temperature tensor (left); time-averaged mass,
momentum, and energy fluxes (right).  The system dimensions are $200 \times 40$ .  The mass
flux of unity is imposed by the rate at which fresh particles are inserted at the left
boundary.
}
\end{figure}

Figure 2 shows a typical Joule-Thomson steady-state particle snapshot, with similar pair and
barrier potentials chosen to minimize integration errors using fourth-order Runge-Kutta
molecular dynamics with a timestep $dt = 0.01$ :
$$
\phi_{\rm pair} (r<1) = [ \ 1 - r^2 \ ]^4 \ ; \ \phi_{\rm barrier} (-1<x<+1) = (1/4)[ \ 1 - x^2 \ ]^4 \ . 
$$
Although such a potential was perfectly satisfactory for the shockwave simulations of twofold
compression it suggests the possibility of poor behavior at high density, where the force is a
decreasing function of compression.  Accordingly we compared results with a modified pair potential
for which the force remains constant, with its maximum value $F_{\rm max}$ at separations less than
$r_{\rm max} = \sqrt{(1/7)} = 0.377964473$ :
$$
\phi_{\rm max} (r<r_{\rm max}) = (6/7)^4 + F_{\rm max}(r - r_{\rm max}) \ ; \
   F_{\rm max} (r<r_{\rm max}) = 8(6/7)^3\sqrt{(1/7)} \ .
$$
Joule-Thomson profiles including this $\phi_{\rm pair}$ precaution weren't significantly changed
from those with the unmodified potential.

Corresponding time-averaged density and velocity profiles are shown in Figure 3, along with the
(necessarily constant) mass flux, $\rho u$ .  Just as in our shock work the one-dimensional
grid-profile averages were all computed using Lucy's one-dimensional smooth-particle weight
function\cite{b7,b8,b9,b10,b11}, with $h = 3$ :
$$
\langle \ f(x_g) \ \rangle = \sum_{x_j>x_g-h}^{x_j<x_g+h}f_jw_{gj} \ ; \
w_{gj} = (5/12)[ \ 1 + |x_{gj}| \ ][ \ 1 - (|x_{gj}|/3) \ ]^3 \rightarrow \int_{-3}^{+3}w(|x|)dx \equiv 1 \ .
$$
With an input speed of 0.5, which quickly accelerates to 0.62, the velocity speeds up to 1.25 on
passing through the plug potential. Straightforward Runge-Kutta simulation converges relatively
simply and quickly to a flow satisfying slight modifications of the conservation relations which
hold for shockwave simulations.  

The longitudinal momentum flux drops at the barrier because the barrier force removes momentum.  The
overall flux drop exactly matches $(F_{\rm barrier}/L_y)$ , with
$$
[ \ P + \rho u^2 \ ]_{\rm left} = [ \ P +\rho u^2 \ ]_{\rm right} - (F/L_y)  \ [ \ F \ {\rm negative} \ ] \ .
$$
Mass, momentum, and energy fluxes are shown in  Figure 3.

The energy flux is particularly interesting.  Adding the contributions of pair interactions
$(\frac{\dot x_i+\dot x_j}{2})x_{ij}F^x_{ij}$ to the ``convective flux''
gives perfect agreement between the entrance and exit flows.  These contributions can be
divided equally between Particles $i$ and $j$.  Alternatively, they can be velocity-weighted:
$(\frac{\dot x_i}{2})x_{ij}F^x_{ij}$ for $i$ and $(\frac{\dot x_j}{2})x_{ij}F^x_{ij}$ for $j$. The
effect of this choice on the energy flux is insignificant, of order $0.001$.  In shockwaves the
{\it total} pressure-tensor component $P_{xx}$ includes the $\rho kT_{xx}$ which is absent in our
Joule-Thomson flux.  The derivations for these two slightly different expressions for
the energy flux are both familiar textbook fare\cite{b15}.  The reason for the difference is
interesting.  The $x$ component of the purely kinetic part of the energy flux (excluding the
contributions from $\phi$ and $F$) involves local sums {\it cubic} in the velocity components.  In the
equilibrium case the cubic sum can be expressed in terms of the stream velocity and the deviations
from it, which can in turn be expressed in terms of temperature :
$$
\langle \  (v_x/2)(v_x^2 + v_y^2) \ \rangle = (1/2)\langle \  (u + \delta v_x)^3 + u(\delta v_y)^2 \ \rangle =
(1/2)u^3 + (3/2)ukT_{xx} + (1/2)ukT_{yy} \ .
$$
The resulting ``extra'' $\rho ukT_{xx}$ can, if desired, be combined with the potential part of
$uP_{xx}$ so as to agree with the continuum energy-flux expression.  Far from equilibrium this
simplification does not hold and the full cubic kinetic-theory sums must be evaluated.

In general, it is interesting to note that the hot and cold momentum fluxes don't match in the
Joule-Kelvin experiment though they do in the shockwave.  (If {\it both} the fluxes, energy and
momentum, were to match, either the shockwave or the throttling experiment would violate the Second Law!)
The reason for the flux drop at the barrier is the latter's contribution to the momentum flux, by
exerting a nonzero compressive force on the hot fluid.  In our demonstration problem the fluid is
cooled substantially, in keeping with the familiar commercial mechanism using throttling as a
model refrigerator.  Like shockwaves, the present high-speed Joule-Thomson
flows are contained by {\it equilibrium} thermodynamic boundaries.  A {\it series} of Joule-Thomson
states can be generated by using several plug barriers rather than just one.  Accordingly, we believe
that they will, like shockwaves, provide a useful source of computer-experimental constitutive
information for flow states far from equilibrium and help in choosing the optimum weight function
for correlating microscopic and macroscopic flow descriptions.

\end{document}